\documentstyle[epsf,bm,graphicx]{aipproc2}
\newcommand{\boldk}{\bm{k}}
\newcommand{\boldn}{\bm{n}}
\newcommand{\boldp}{\bm{p}}
\newcommand{\boldx}{\bm{x}}
\newcommand{\boldI}{\bm{I}}
\newcommand{\boldtheta}{\bm{\theta}}
\newcommand{\boldomega}{\bm{\omega}}
\newcommand{\boldalpha}{\bm{\alpha}}
\def\<{\langle}
\def\>{\rangle}
\newcommand{\tfrac}[2]{{\textstyle\frac{#1}{#2}}}
\newcommand{\enumparams}{%
 \renewcommand{\labelenumi}{\arabic{enumi})~}
 \setlength{\itemindent}{1.5em}
 \setlength{\parskip}{0pt}
}

\begin{document}

\title{Periodic Orbits and Deformed Shell Structure
\thanks{
Talk presented by K.M. at the Conference on 
{\it Frontiers of Nuclear Structure},
July 29th - August 2nd, 2002, UC Berkeley.}
}

\author{
 Ken-ichiro Arita$^{\sl a}$,
 Alexander G. Magner$^{\sl b}$ 
and
 Kenichi Matsuyanagi$^{\sl c}$}
\address{%
 $^{\sl a}$Department of Physics, Nagoya Institute of Technology,
 Nagoya 466-8555, Japan\\
 $^{\sl b}$Institute for Nuclear Research, 03680 Prospekt Nauki 47,
 Kiev-28, Ukraine\\
 $^{\sl c}$Department of Physics, Graduate School of Science,
 Kyoto University,\\ Kitashirakawa, Kyoto 606-8502, Japan}

\maketitle

\begin{abstract}
Relationship between quantum shell structure and classical periodic
orbits is briefly reviewed on the basis of semi-classical 
trace formula. Using the spheroidal cavity model, it is shown that
three-dimensional periodic orbits, which are born out of bifurcation 
of planar orbits at large prolate deformations, 
generate the superdeformed shell structure.   
\end{abstract}

\subsection*{Introduction}

\begin{figure}[t]
\begin{center}
\includegraphics[width=.48\textwidth]{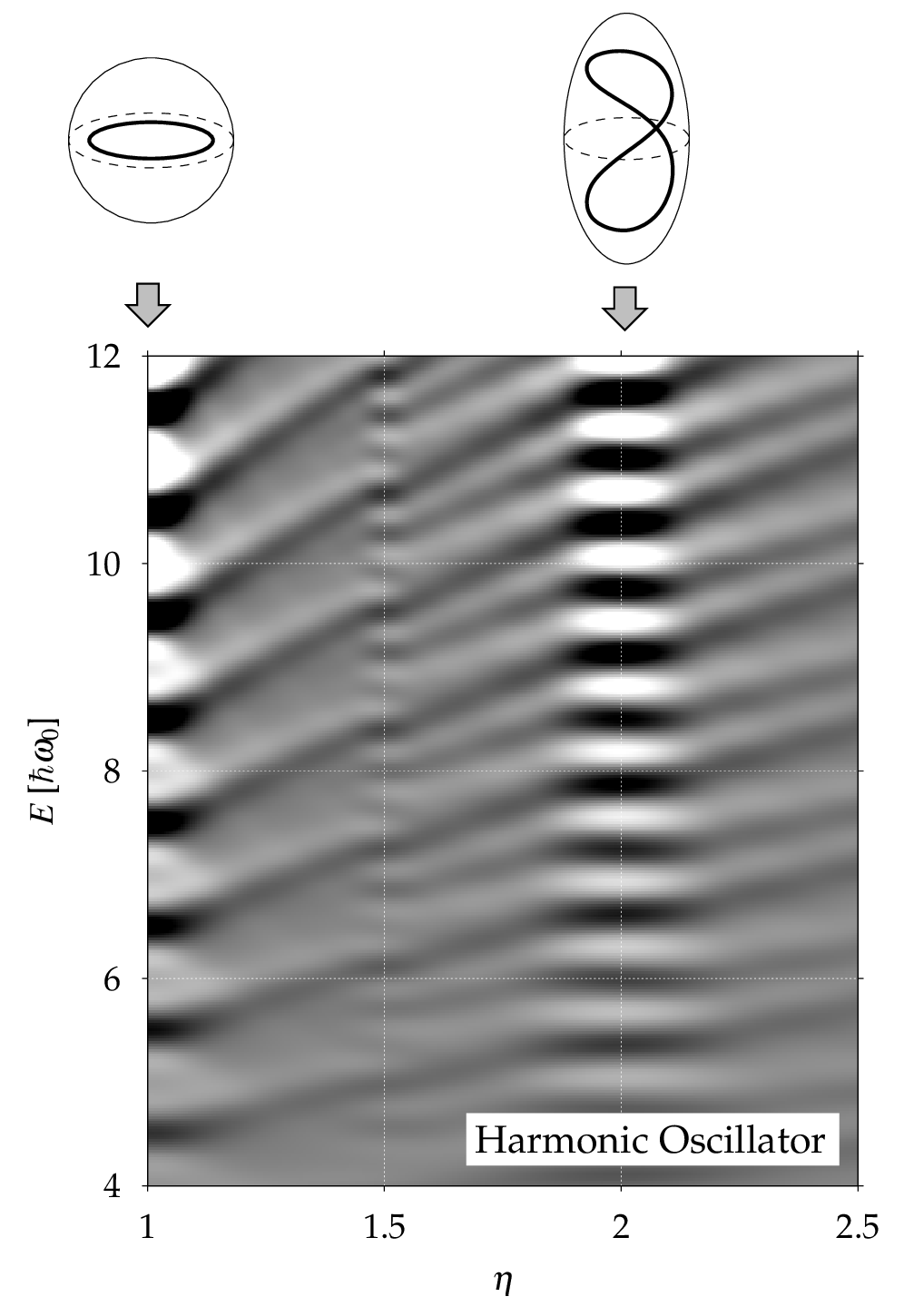}
\end{center}
\caption{\label{fig1}
Oscillating level density as a function of axis ratio $\eta$
for the axially symmetric HO potential. 
Bright and dark regions correspond to high and low level densities,
respectively.
The 3D periodic orbits exist only at the 2:1 shape.}
\end{figure}

Existence of superdeformed (SD) nuclei is often explained in terms of
the SD magic numbers for the harmonic-oscillator (HO) potential with
axis ratio 2:1.  It appears, however, that we need a more general
explanation not restricted to the HO potential, since, up to now, more
than 200 SD bands have been found in various regions of nuclear chart
and their shapes in general deviates from the 2:1 shape to some
extent.  In this talk, we shall discuss the mechanism how and the
reason why the SD shell structure emerges.  The major tool for this
purpose is the trace formula, which is the central formula in the
semiclassical periodic-orbit (PO) theory and provides a link between
quantum shell structure and classical periodic orbits in the mean
field.  Here, shell structure is defined as regular oscillation in the
single-particle level density coarse-grained to a certain energy
resolution.  An example of coarse-graining for the well-known axially
symmetric HO model is displayed in Fig.~\ref{fig1}.

In this talk, we discuss the spheroidal cavity model, since, in
contrast to the HO model, this model is very rich in periodic orbits;
it is an ideal model for exhibiting the presence of various kinds of
periodic orbit and their bifurcations.  We present both Fourier
transforms of quantum spectra and semiclassical calculations based on
the PO theory, and identify classical periodic orbits responsible for
emergence of the SD shell structure.  The result clearly shows that
three-dimensional (3D) periodic orbits, that are absent in spherical
and normal deformed systems and are born out of bifurcations of planar
orbits, generate a new shell structure at large prolate deformations,
which may be called ``the SD shell structure.''  They continue to
exist for a wide range of deformation, once they are born.

The PO theory provides a basic tool to get a deeper understanding of
microscopic origin of symmetry breaking in the mean field.  It sheds
light, in addition to the stability of the SD nuclei, on the reason of
prolate dominance in normal deformed nuclei, on the origin of
left-right asymmetric shapes, etc.  It is useful for finite
many-Fermion systems covering such different areas as nuclei, metallic
clusters, quantum dots, etc. In this talk, we shall also touch upon
such applications of the PO theory.

\subsection*{Level Bunching and Trace Formula}

For the axially symmetric HO potential, the following two conditions
coincide:
\begin{enumerate}\enumparams
\item Axis ratio~~~ $\< x^2 \>:\< z^2 \> = 1:2$
\item Frequency ratio~~~ $\omega_x:\omega_z = 2:1\,,$
\end{enumerate}
but they are different in general.  
Condition 2) is nothing but the PO condition, and
possesses a more general significance than condition 1).
We can examine this point as follows.
For any integrable Hamiltonian system, 
we can introduce action and angle variables $(\boldI, \boldtheta)$
which satisfy the canonical equations of motion,
\begin{eqnarray}
\dot{\boldtheta} = \frac {\partial H}{\partial \boldI} =\boldomega(\boldI),
\end{eqnarray}
and the energy $E$ can be quantized by 
the EBK (Einstein-Brillouin-Keller) quantization condition:
\begin{eqnarray}
E_{\boldn} = H(\boldI)~~~~~{\rm with}~~~ 
\boldI = \hbar\,(\boldn + \tfrac14\boldalpha),
\end{eqnarray}
where $\boldn$ represents a set of quantum numbers, 
$\boldn=(n_1,n_2,n_3)$ with $n_i=0,1,2,\cdots$,
and $\boldalpha$ the Maslov indices.
We see that level degeneracy occurs when
\begin{eqnarray}
E_{\boldn + \Delta \boldn} - E_{\boldn}
 &=&H(\boldI + \Delta \boldI) - H(\boldI)
\nonumber\\
 &\simeq&\frac {\partial H}{\partial \boldI}\Delta \boldI
\nonumber\\
 &=& \hbar \boldomega \cdot \Delta \boldn
\nonumber\\
&=&\hbar\omega_1\Delta n_1+\hbar\omega_2\Delta n_2 +\hbar\omega_3\Delta n_3
\nonumber\\
&=&0,
\end{eqnarray}
i.e., when $\omega_1:\omega_2:\omega_3$ are in rational ratios.  This is
just the condition for the classical orbit to be periodic, and
discussed in detail in the textbook of Bohr and Mottelson\cite{bor75}.

On the basis of the semiclassical PO theory, we can examine, in a more
general way, the decisive role of periodic orbits as origin of level
bunching. According to this theory (see, e.g., \cite{bra97a} for a
review), the level density $g(E)$ is given by a sum of the average
part $\bar g(E)$ and the oscillation part $\delta g(E)$ as
\begin{eqnarray}
  g(E)&=&\sum_n\delta(E-E_n) \nonumber\\
     &\simeq& \bar g(E)+ \delta g(E) \nonumber\\
     &=& \bar g(E)+
     \sum_\alpha A_\alpha \cos\left(\frac{1}{\hbar}S_\alpha(E)
     -\frac{\pi}{2}\mu_\alpha\right),
\end{eqnarray}
where $S_\alpha(E)$ denotes the action of the periodic orbit $\alpha$,
and $\mu_\alpha$ is a phase related with the Maslov index.  This
equation is called ``trace formula'' and provides a link between
quantum shell structure and classical periodic orbits in the mean
field.  There is a complementarity between the energy resolution and
periods of classical orbits, so that, for the purpose of understanding
the origin of regular oscillation patterns in the smoothed
single-particle level density (shell structure), we need only short
orbits in the sum over $\alpha$.

\begin{figure}[b]
\noindent
\begin{minipage}{.48\textwidth}
\noindent
\includegraphics[width=\textwidth]{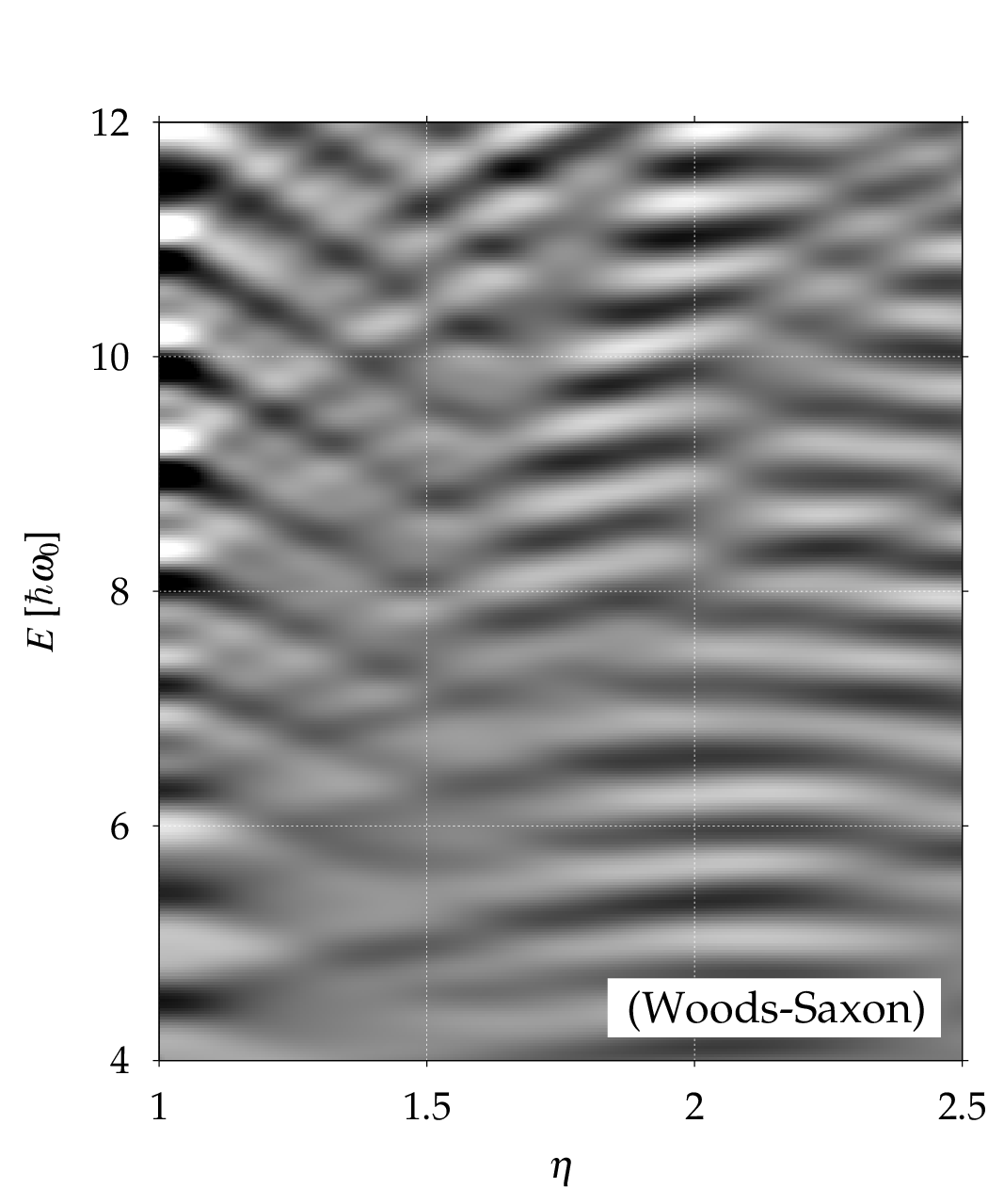}
\caption{\label{fig2}
Oscillating level density as a function of axis ratio $\eta$
for the WS potential without the spin-orbit term.
Bright and dark regions correspond to high and low level densities,
respectively.
The calculation was done following the procedure described on p.~593
of Ref.~\protect\cite{bor75}.}
\end{minipage}
\hfill
\begin{minipage}{.48\textwidth}
\noindent
\includegraphics[width=\textwidth]{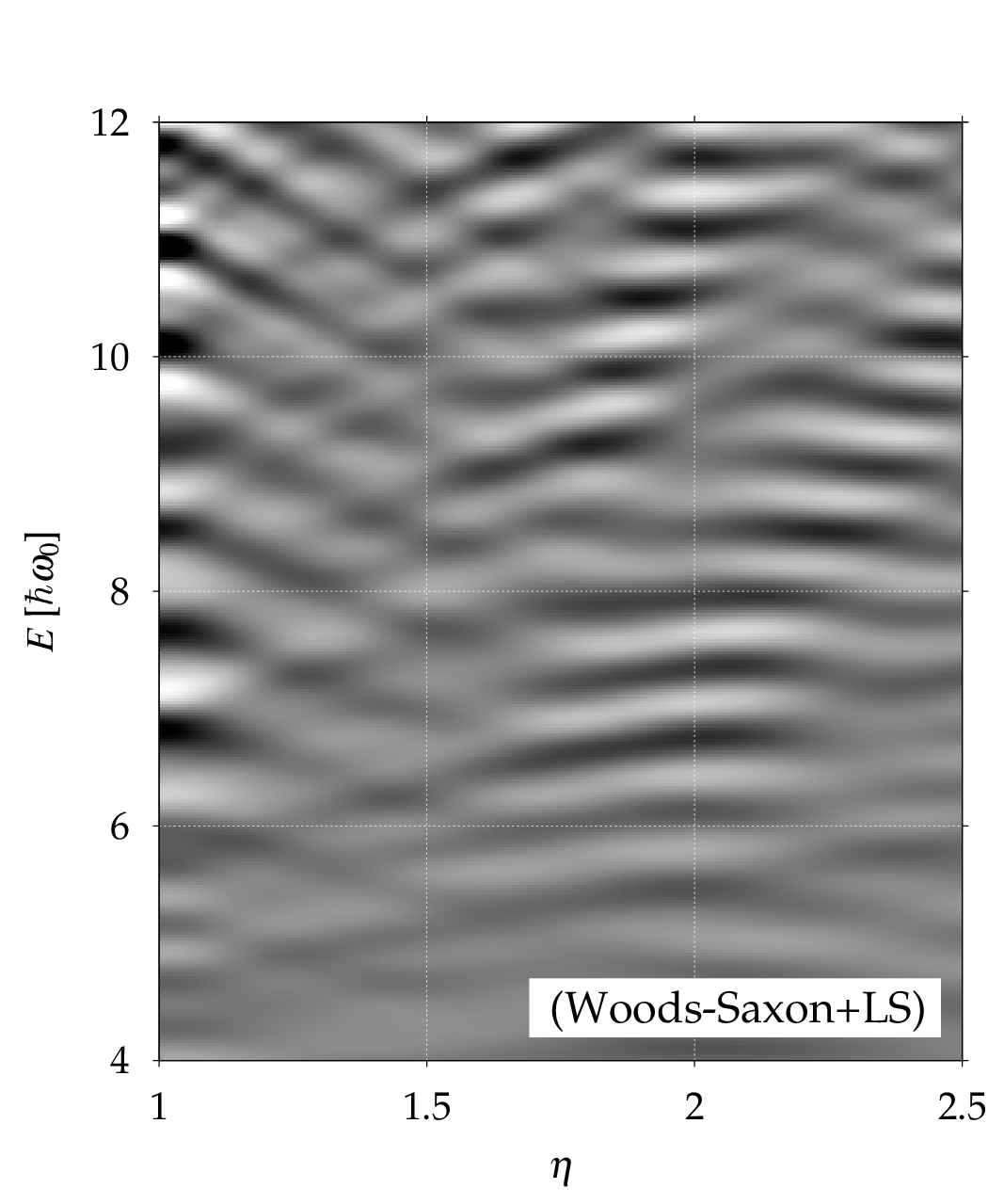}
\caption{\label{fig3}
Oscillating level density as a function of axis ratio $\eta$ 
for the WS potential.
Bright and dark regions correspond to high and low level densities,
respectively.
Here, the spin-orbit term with 
$v_{ls} = -0.12$ is added to the Hamiltonian
used in Fig.~\protect\ref{fig2}.}
\end{minipage}
\end{figure}

\subsection*{Periodic Orbits and Shell Structure in the Cavity Model}

\begin{figure}[t]
\begin{center}
\includegraphics[width=.9\textwidth]{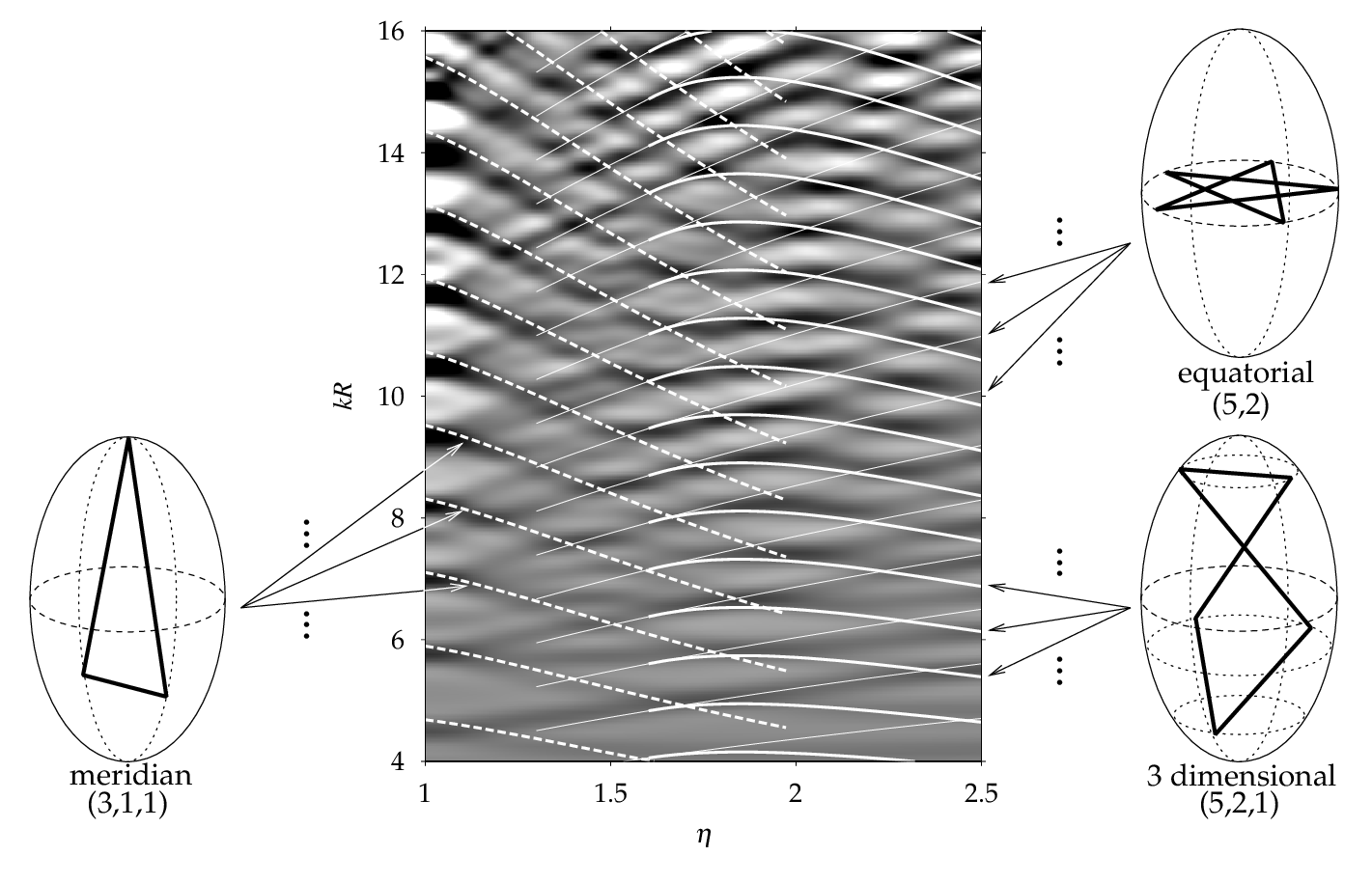}
\end{center}
\caption{\label{fig4}
Ridge-valley structure in the oscillating level density for the
spheroidal cavity model.  Bright and dark regions correspond to high
(ridge) and low (valley) values of the oscillating level density,
respectively.  Constant action lines running along valleys are shown
for typical periodic orbits by dashed, thin-solid, and thick-solid
curves.  Note that they are indicated as representatives of the
meridian-plane orbits, the equatorial-plane orbits, and the 3D orbits,
respectively: There are several families of periodic orbits with
similar lengths, and constant-action lines of them also behave in the
same way as those shown here, see Refs.~\protect\cite{str77} and
\protect\cite{ari98} for details.}
\end{figure}

Let us consider the cavity model, which may be regarded as a
simplified model of Woods-Saxon potential for heavy nuclei.  In fact
their basic patterns of shell structure are similar with each
other. Certainly, the spin-orbit term shifts the magic numbers, but it
does not destroy the valley-ridge structure discussed below.  One can
confirm these points by comparing Figs.~\ref{fig2}, \ref{fig3} and
\ref{fig4}.
 
Role of periodic orbits for shell structure in this model was
originally studied by Balian and Bloch\cite{bal72}, and has been
discussed in the investigations of
\begin{enumerate}\enumparams
\item the reason of prolate dominance in nuclear shape
\cite{str77,arv87,fri90}
\item the supershell effects in metallic clusters\cite{nis90}
\item the origin of left-right asymmetric shapes\cite{bra97b,sug98},
~~etc.
\end{enumerate}
 
For cavity models, the energy and the momentum are simply related as 
\begin{eqnarray}
  E=\boldp^2/2m, \qquad \boldp=\hbar\boldk,
\end{eqnarray}
and the action integral is proportional to the length $L_\alpha$,
\begin{eqnarray}
  S_\alpha=\oint_\alpha \boldp\cdot d\boldx
  = \hbar kL_\alpha.
\end{eqnarray}

Accordingly, the trace formula for the level density can be written as
\begin{eqnarray}
  g(k) \simeq \bar g(k)+
     \sum_\alpha A_\alpha \cos\left(kL_\alpha -\frac{\pi}{2}\mu_\alpha\right).
\end{eqnarray}
It can be easily confirmed that only short orbits contribute to the
level density coarse-grained in energy.  Let us Fourier transform the
level density
\begin{eqnarray}
F(L)&=&\int dk e^{ikL} g(k) \nonumber\\
    &\simeq& \sum_\alpha \tilde A_{\alpha} \delta(L-L_\alpha).
\end{eqnarray}
This equation indicates that peaks will show up at lengths $L_\alpha$
of periodic orbits $\alpha$, which may be called ``length spectrum''.
Now, the orbit lengths change when the deformation parameter $\eta$
varies.  Let us then consider the oscillating level density as a
function of $\eta$,
\begin{eqnarray}
\delta g(k,\eta)
\simeq \sum_\alpha A_\alpha(k,\eta) 
\cos\left(kL_\alpha(\eta) -\frac{\pi}{2}\mu_\alpha\right).
\end{eqnarray}
From this formula, we see that, if a few orbits dominate in the sum,
the valley-ridge structure on the $(k,\eta)$ plane will be determined
by the constant action lines,
\begin{eqnarray}
kL_\alpha(\eta)={\rm const},
\end{eqnarray}
of these dominant orbits\cite{str77,ari98}.  In fact, we see in
Fig.~\ref{fig4} that the valley-ridge structure is well explained in
terms of three kinds of short periodic orbit.
\begin{figure}[b]
\begin{center}
\includegraphics[width=.9\textwidth]{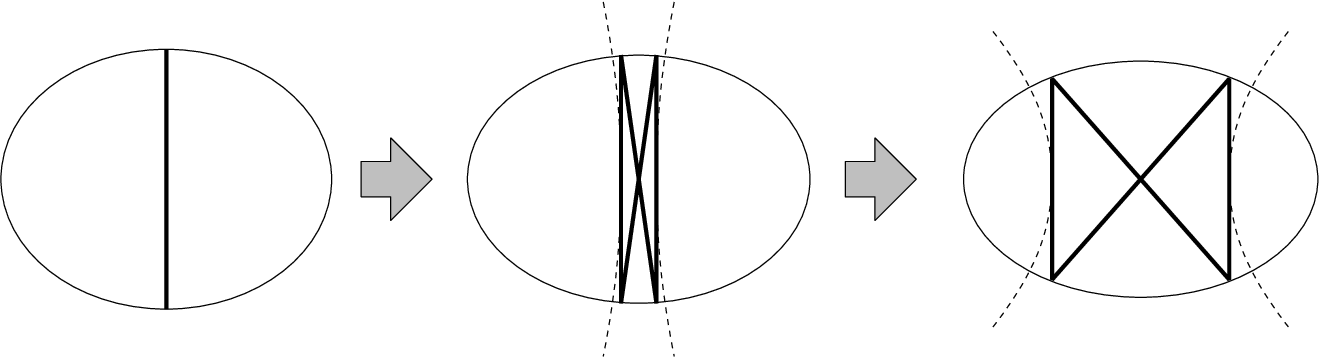}
\end{center}
\caption{\label{fig5}
Birth of a butterfly shaped orbit from the short diameter
through bifurcation at $\eta=\sqrt{2}$ on the meridian plane.
This figure illustrates a representative orbit among a continuous
family of orbits possessing the same length.
}
\begin{center}
\includegraphics[width=.75\textwidth]{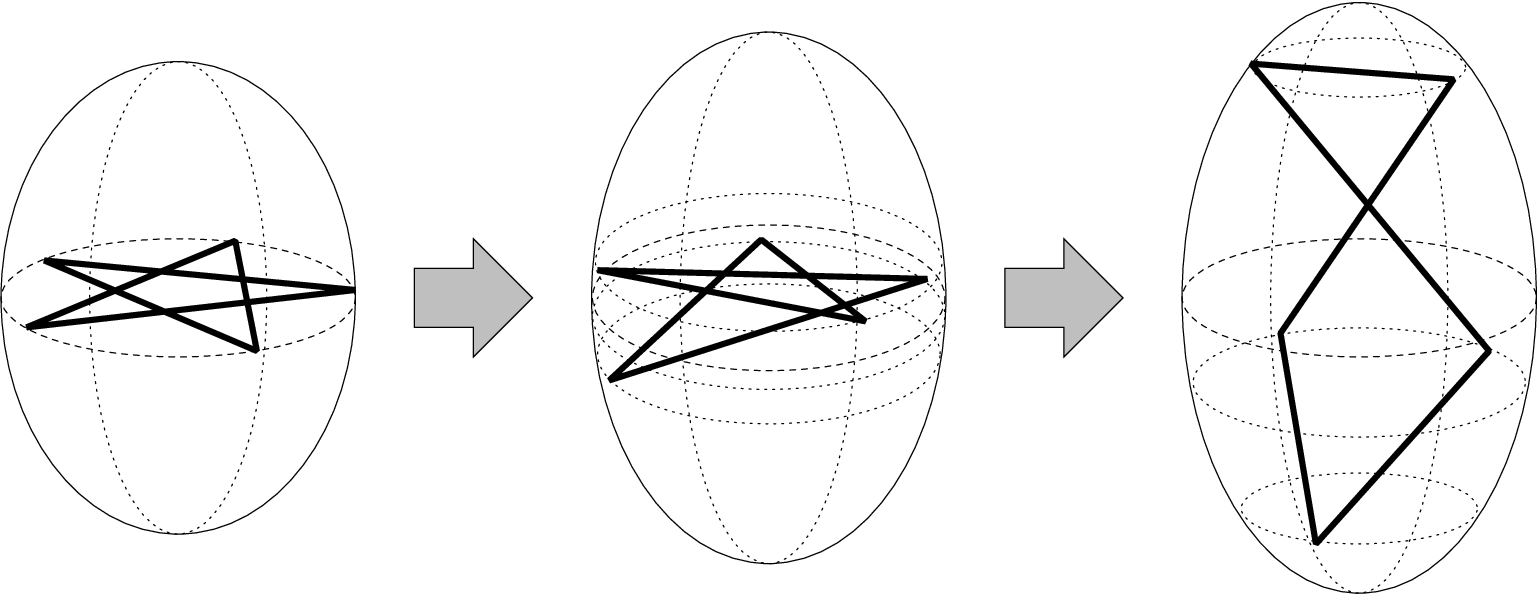}
\end{center}
\caption{\label{fig6}
Birth of a 3D orbit from the star-shaped orbit on the equatorial plane 
through bifurcation at $\eta \simeq 1.62$.
This figure illustrates a representative orbit among a continuous
family of orbits possessing the same length.
}
\end{figure}

\begin{figure}[t]
\noindent
\begin{minipage}{.48\textwidth}
\includegraphics[width=\textwidth]{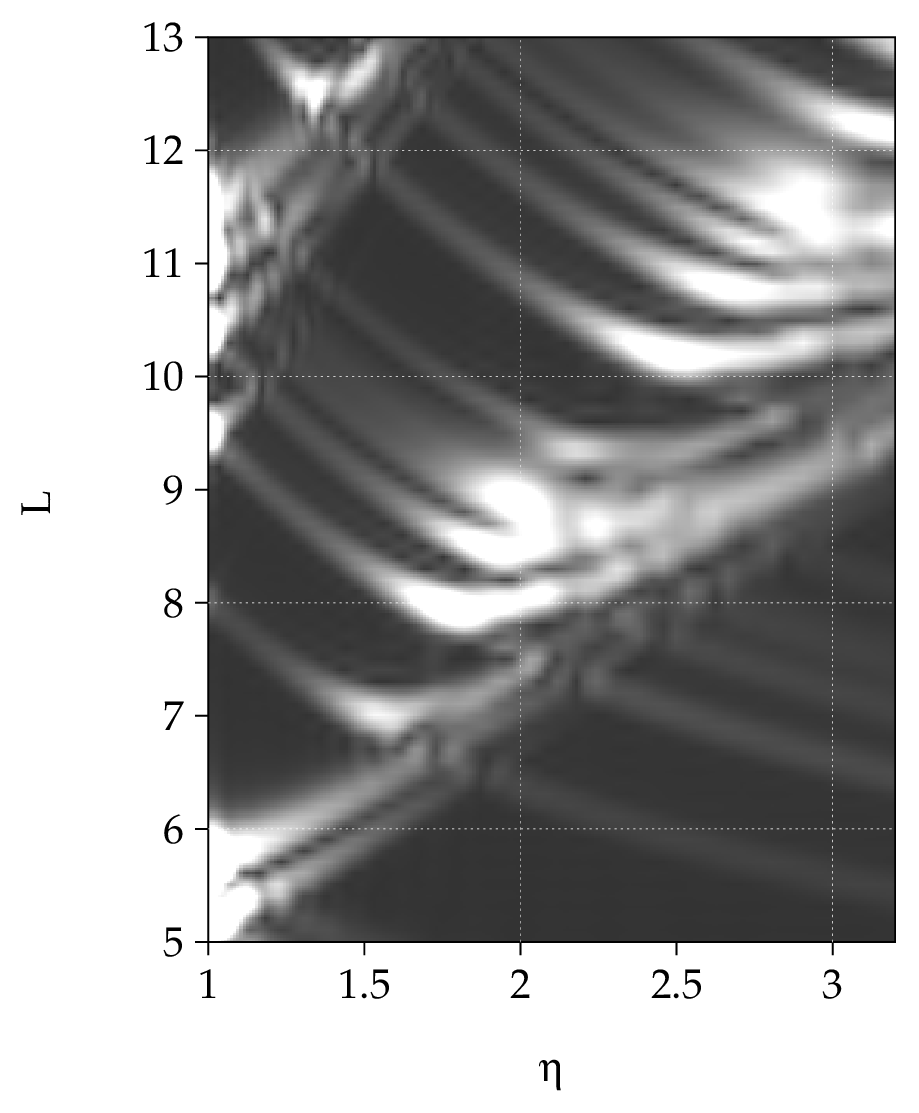}
\end{minipage}
\hfill
\begin{minipage}{.48\textwidth}
\includegraphics[width=\textwidth]{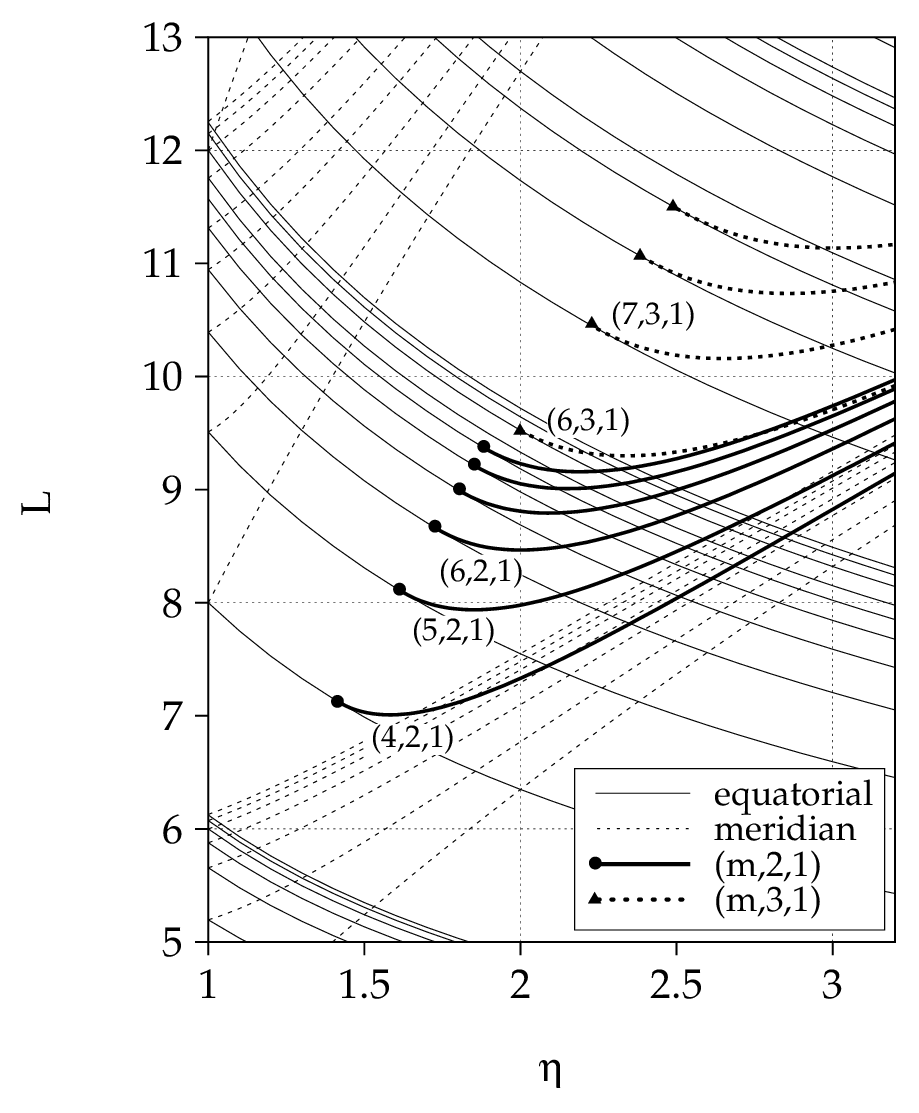}
\end{minipage}
\caption{\label{fig7}
Left-hand part: Absolute values of the Fourier amplitudes,
$|F(L,\eta)|$, of the level density, drawn as a map on the $(L,\eta)$
plane.  Brightness is proportional to the magnitude. Right-hand part:
Lengths of various classical orbits as functions of axis ratio $\eta$.
The 3D orbits responsible for the SD shell structure are denoted as
(4,2,1), (5,2,1), (6,2,1), etc., and their lengths are plotted by
thick-solid curves, while those of the equatorial-plane orbits and of
the meridian-plane orbits are plotted by thin-solid and dashed
curves, respectively.@The bifurcation points of the 3D orbits of
interest are indicated by filled circles.  (Filled triangles indicate
bifurcation points of 3D orbits responsible for hyperdeformed shell
structure not discussed here.)  See Refs.~\protect\cite{ari98} and
\protect\cite{mag02} for details.}
\end{figure}

\begin{figure}[t]
\begin{center}
\includegraphics[width=.6\textwidth]{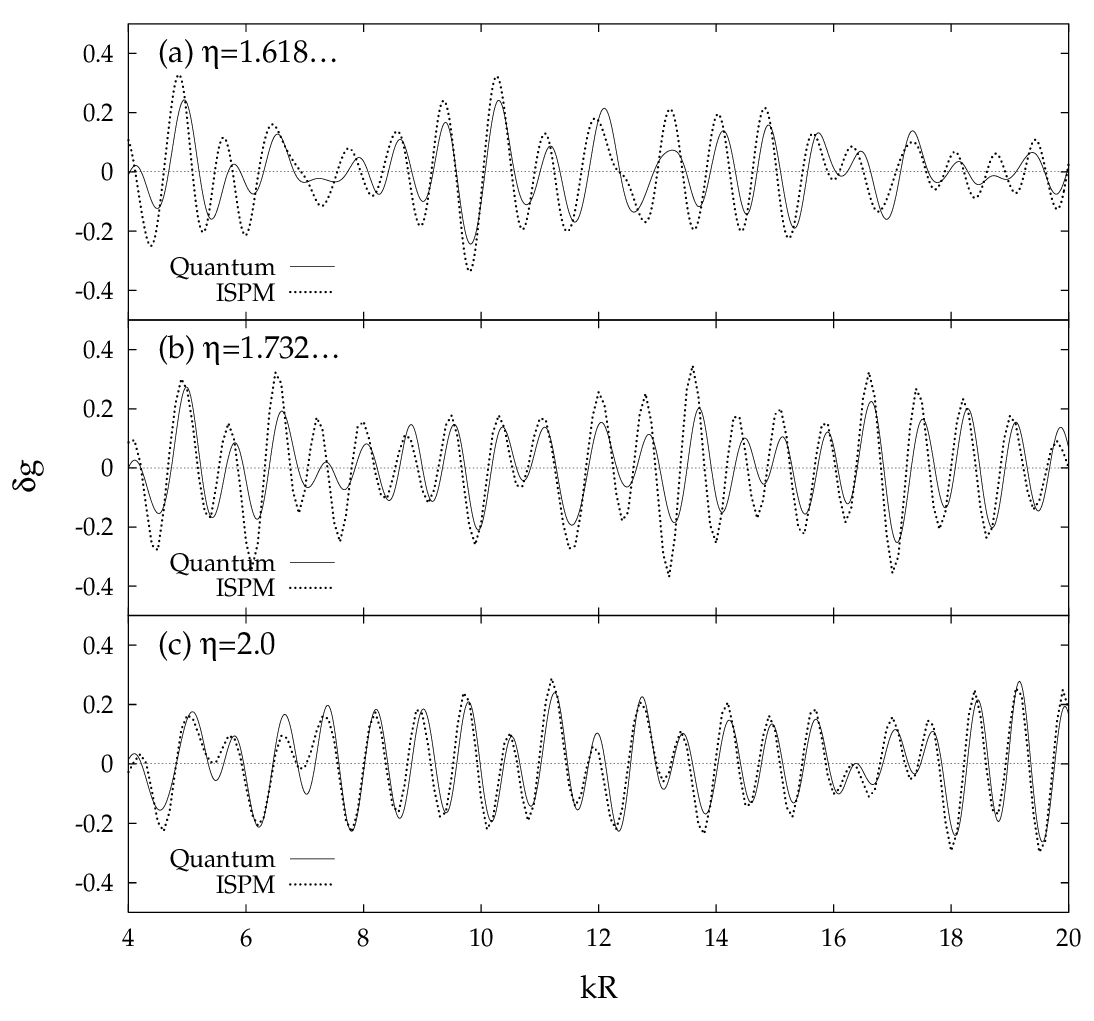}
\end{center}
\caption{\label{fig8}
Comparison between the oscillating level densities evaluated by the
semiclassical trace formula (dotted lines denoted ISPM) and those
obtained by quantum mechanics (solid lines denoted Quantum).  They are
plotted for several values of $\eta$ as function of $kR$, where $R$
denotes the radius of the cavity in the spherical limit (the volume of
cavity is conserved during variation of $\eta$).}
\end{figure}

\subsection*{Bifurcations}

As illustrated in Figs.~\ref{fig5} and \ref{fig6}, when the axis ratio
$\eta$ of the spheroidal cavity reaches $\sqrt{2}$, the
butterfly-shaped planar orbits emerge on the meridian plane through
bifurcations of the linear orbit along the short diameter.  When
$\eta$ further increases, 3D orbits emerge at $\eta \simeq 1.62$
through bifurcation of the five-point star shaped orbits on the
equatorial plane. Likewise, other 3D orbits appear at $\eta=\sqrt{3}$
through bifurcation of second repetitions of the triangular orbits on
the equatorial plane, \ldots, etc.  Note that the figure illustrates
representative orbits only. In fact, each $\alpha$ in the trace
formula (9) represents a continuous family of orbits with the same
topology possessing the same values of action (length).  In contrast
to the HO potential, these 3D orbits continue to exist, once they
appear through the bifurcations.

Peaks in the Fourier transform (8) of the level density
will follow the variations of orbit lengths $L_\alpha$ with $\eta$.
Thus, we can draw a map of the Fourier amplitudes on the $(L,\eta)$ plane,
\begin{eqnarray}
F(L,\eta)=\sum_\alpha \tilde A_{\alpha} \delta(L-L_\alpha(\eta)).
\end{eqnarray}
In Fig.~\ref{fig7}, the Fourier amplitudes are compared with lengths
of classical periodic orbits.  This figure exhibits a beautiful
quantum-classical correspondence.  Furthermore, by comparing the
bright regions in the left-side figure with the bifurcation points
indicated in the right-side figure, we find significant enhancement of
the shell structure amplitudes just on the right-hand side of the
bifurcation points.

Unfortunately, the amplitude $A_{\alpha}(k,\eta)$ in the trace formula
(9) diverges at the critical point of deformation $\eta$ where the
orbit bifurcation takes place.  This is because the stationary phase
approximation used in the standard semiclassical PO theory breaks down
there.  Thus, the standard trace formula is unable to describe the
enhancement phenomena seen in Fig.~\ref{fig7}.  To overcome this
difficulty, in recent years, we have developed a new semiclassical
approximation scheme, called an improved stationary phase
approximation, and derived a new trace formula free from such
divergence\cite{mag99,mag01,mag02}.  A numerical example obtained by
this approach is presented in Fig.~\ref{fig8}.  We see that the basic
pattern of oscillation in the quantum level density at large
deformation is nicely reproduced by the semiclassical calculation
using the new trace formula.  In this way, we have
confirmed\cite{ari98,mag01,mag02} that, in the region of large prolate
deformation with axis ratio $\eta \geq 1.62 $ (which corresponds to
the ordinary deformation parameter $\delta \geq 0.44$), the major
pattern of the oscillating level density is determined by
contributions from the bifurcated 3D orbits.

\subsection*{Conclusion}

The 3D periodic orbits generate a new shell structure at large prolate
deformations. We may call this shell structure ``SD shell structure.''
These 3D orbits are born out of bifurcations of planar orbits in the
equatorial plane, and they play dominant roles immediately after the
bifurcations.  Thus, the SD shell structure is a beautiful example of
emergence of new structure through bifurcation, and may be regarded as
quantum manifestation of classical bifurcation.

\end{document}